# Experimental signature of bandgap opening in bilayer graphene at metal contacts

Ryo Nouchi[a]

Nanoscience and Nanotechnology Research Center, Osaka Prefecture University, Sakai, Osaka 599-8570, Japan

[a] Electronic mail: r-nouchi@21c.osakafu-u.ac.jp

**ABSTRACT:** Bilayer graphene (BLG) possesses a finite bandgap when a potential difference is introduced between the two graphene layers. The potential difference is known to be introduced by surface charge transfer. Thus, it is expected that a finite bandgap exists at the metal contacts. The bandgap at the metal-BLG interface can be detected by the superlinear current-voltage characteristics in back-gate field-effect transistors, caused by carriers tunneling through the bandgap. The superlinearity was higher in the positively-gated region, attributed to hole doping from the Cr/Au electrodes. The control experiments using single-layer graphene (SLG) did not have a superlinearity, which is consistent with the fact that a sizeable bandgap is not expected at the metal-SLG interface. The opening of a bandgap at the metal-BLG interface is an additional source of electrode-contact resistance.

Graphene has a high charge carrier mobility[1] and is expected to be an ideal channel material in electronic devices that operate at high speeds. However, single-layer graphene (SLG) possesses no bandgap, hindering its usefulness in a wide range of applications such as digital logic circuits. The introduction of a bandgap in SLG and multilayered analogs has been a subject of intense research. Efforts include quantum confinement by forming nanoribbons[2–4] and applying a perpendicular electric field to bilayer graphene (BLG).[5–8] The former technique is suitable for high-density integration because of the ultra-narrow width of the channel. The bandgap of the materials is determined by its width,[2] but microscopic fluctuations in the width generally exist in lithographically patterned nanoribbons,[9] leading to an inhomogeneous band profile and a reduction of the charge carrier mobility.[10] The latter technique offers bandgap controllability (determined by the magnitude of the potential difference between two graphene layers) and does not require precise lithographical techniques on the nanometer scale. The potential difference can be generated by applying a gate voltage[6–8] and by introducing a difference in the carrier concentration between the two graphene layers by means of surface charge transfer from the adsorbed layers.[11,12]

The most common structure of a BLG field-effect transistor (FET) has a double gate configuration. This enables the strength of the displacement field and the carrier concentration in the BLG to be controlled independently. Although it is unsuitable to control these two parameters independently, opening the bandgap by surface charge transfer is based on a single gate (back-gate) device, which has a simple device configuration. These two techniques require different device configurations. However, these two mechanisms (a gate electric field and surface charge transfer) can coexist in both device configurations. In back-gate devices, applying a gate voltage to the back-gate electrode generates a vertical electric field, although the strength of the resultant displacement field between the two constituent graphene layers is not as strong as that in the double-gate configuration. In double-gate devices, in addition to foreign molecules/atoms unintentionally adsorbed on the surface, the metal contacts induce charge transfer, caused by the Fermi level alignment between the metallic electrodes and the BLG. While the number of surface adsorbates is expected to be reduced by improving the device fabrication processes, metal contacts are a required component in FETs. However, the bandgap opening in BLG at the metal contacts has not been addressed experimentally thus far.

In this Letter, we experimentally examine the bandgap of BLG at the metal contacts in single-gate (back-gate) FETs. Superlinear electric-current (the drain current $I_D$) vs. voltage (the drain voltage $V_D$) characteristics were observed in back-gate BLG FETs. The superlinearity could be tuned by the gate voltage ($V_G$), and was higher in the positively-gated regions. This result is consistent with hole doping from the metal contacts, which forms a *pnp* junction in the positively-gated region and causes a current to tunnel through bandgap at the two *pn* junctions.

The graphene flakes were fabricated by mechanical exfoliation from a graphite crystal using adhesive tape. They were placed onto a highly-doped Si substrate with a 300-nm-thick thermal oxide layer on top of it. The doped Si and $SiO_2$ layers were the gate electrode and gate dielectric, respectively. The number of layers in the graphene flakes was confirmed by the optical contrast and Raman scattering spectra. Source/drain electrodes were patterned by electron-beam lithography and were formed by thermally evaporating metals followed by a liftoff process. The electrodes were Au films with a 0.7-nm-thick Cr adhesion layer. The electrode contact length along the channel length direction was 1 μm. After liftoff, the fabricated back-gate FETs were rinsed in acetone and 2-propanol. The electrical measurements were performed in air in the dark and at room temperature unless otherwise stated.

Figure 1a shows the transfer ($I_D$–$V_G$) characteristics for an as-fabricated BLG FET with a channel length, $L = 0.5$ μm and $V_D = 10$ mV. The non-zero and positive charge-neutrality point ($V_{NP}$, defined as the $V_G$ that corresponds to the minimum $I_D$) indicate the presence of charged impurities that doped the BLG with holes and are likely environmental molecules such as oxygen and water.[13,14] The output ($I_D$–$V_D$) characteristics for the same FET were also measured with $V_G$ ranging from $V_{NP} - 60$ to $V_{NP} + 60$ V in 10 V steps (see Fig. 1b). Superlinear curves are clearly observed. To parameterize the nonlinearity, the differential conductances ($dI_D/dV_D$) were calculated from the output characteristics in Fig. 1b using a difference method (Fig. 1c). For the superlinear curves, the difference calculated by subtracting the zero-bias conductance from the $dI_D/dV_D$ value at 1 V should be positive, while it should be negative for the sublinear curves. This difference, $\Delta(dI_D/dV_D)$, is plotted against $V_G - V_{NP}$ in Fig. 1d. The degree of the nonlinearity had a $V_G$ dependency, where $\Delta(dI_D/dV_D)$ was positive in the positively-gated region and significantly decreased as $|V_G|$ increased in the negatively-gated region.

There are two types of nonlinearity: namely, sublinearity and superlinearity. Thus, the causes for the nonlinear $I_D$-$V_D$ characteristics are also classified into these two types. Carrier velocity saturation by optical phonon scattering[15,16] and self-heating[17] are known to lead to sublinearities. The charge transport across a charge-injection barrier (Schottky barrier) can be generally characterized by thermionic emission, which is known to cause superlinearities.[18] The transport through a tunneling barrier is also known to be superlinear at relatively high bias voltages.[19] In addition, superlinear-like kinked $I_D$-$V_D$ curves are caused by the formation of an ambipolar channel.[15] Although the actual shape of the $I_D$-$V_D$ curves is determined by the combination of these effects, the observation of superlinear curves unambiguously indicates the presence of the cause(s) for superlinearities. Each cause will be examined below.

An ambipolar channel is formed when the magnitude of $V_D$ is higher than that of $V_G - V_{NP}$. The lowered gate-to-channel potential caused by the finite $V_D$ can induce carriers with the opposite polarity at the drain end. This leads to a sudden increase in the electrical current.[15] As shown in Fig. 1d, superlinear $I_D$-$V_D$ curves are also observed when $|V_G - V_{NP}|$ is higher than the maximum $|V_D|$ used (1 V). In addition, the observed $I_D$-$V_D$ curves had no kink, excluding the possibility of the formation of an ambipolar channel.

One possible source for the tunneling is an unintentionally-formed interfacial layer at the metal-BLG contacts. Possible origins are metal oxide and contamination layers. These thin interfacial layers are known to induce distorted (double-dip-type) transfer characteristics in SLG FETs,[20,21] but this is not the case in the BLG FETs in this study (Fig. 1a). To further confirm the nonexistence of the interfacial layers, SLG FETs were investigated. The formation of a sizable bandgap was not expected at the incommensurate metal-SLG contacts.[22] Thus, a superlinearity caused by tunneling should not be observed in the SLG FETs with no interfacial layer. Figure 1e shows the transfer characteristics of a SLG FET with $L = 0.5$ μm and $V_D = 10$ mV. The output characteristics were measured with $V_G$ ranging from $V_{NP}$ +50 to $V_{NP}$ −50 V in 10 V steps (Fig. 1f), which were used to calculate the $dI_D/dV_D$ and $\Delta(dI_D/dV_D)$ in Figs. 1g and h, respectively. As expected, a superlinearity, i.e., a positive $\Delta(dI_D/dV_D)$, is not observed in Fig. 1h. Instead, negative $\Delta(dI_D/dV_D)$ values are observed, indicating sublinear $I_D$–$V_D$ characteristics. This was a result of carrier velocity saturation by optical phonon scattering[15,16] and/or self-heating.[17] Negative values were also observed in the BLG FETs with large negative $V_G$ (see Fig. 1d). The absence of the superlinearity in SLG FETs was confirmed in 14 SLG devices among the 16 devices tested, while all of the tested BLG FETs (19 devices) showed the superlinearity. This strongly supports that the superlinear output characteristics observed in the BLG FETs originated mostly from sources other than the unintentional formation of interfacial layers at the metal contacts.

A charge-injection barrier generally occurs at a metal contact on a semiconductor with a bandgap. For metal-BLG interfaces, the magnitude of the induced bandgap is predicted to be small (~0.1 eV at Au-BLG contacts[23]). If the Fermi level ($E_F$) of the BLG under the contacts has a larger shift from the charge neutrality point than half of the predicted bandgap, then the BLG under the contact should be degenerately doped. In this case, a charge-injection barrier can be eliminated. Figure 2a shows the $L$-dependence of the transfer characteristics of as-fabricated BLG FETs measured with $V_D = 10$ mV. The $V_{NP}$ shifted to more positive $V_G$ as $L$ became shorter (Fig. 2b), which is attributed to hole doping from the metal contacts.[24] The obtained $V_{NP}$ values are considered to mainly reflect the charge neutrality point of the channel center,[24] because the ability of $V_G$ to control the potential of the BLG under the contacts should be limited (called charge-density pinning), which is the case for SLG FETs.[25] The $E_F$ shift ($\Delta E_F$) was calculated from the $V_{NP}$ values with the relationship, $\Delta E_F = \hbar^2 \pi \varepsilon_0 \varepsilon_r V_{NP}/(2dqm^*)$, where $\hbar$ is the Dirac constant, $\varepsilon_0$ is the permittivity of vacuum, $\varepsilon_r$ is the relative permittivity of the gate insulator (3.9), $d$ is the thickness of the gate insulator (300 nm), $q$ is the unit electronic charge, and $m^* = 0.033m_0$ ($m_0$ is the bare electron mass) is the effective mass of the carriers in BLG.[26] By extrapolating to zero $L$, the $\Delta E_F$ at the contacts was estimated to be 0.25 eV. This corresponds to a hole concentration of $6.9 \times 10^{12}$ cm$^{-2}$, and thus high-energy (valence) bands can be excluded here.[27] The $\Delta E_F$ is sufficiently higher than half of the expected bandgap at the metal-BLG contacts.[23] Thus, the BLG under the contacts should be degenerately doped, justifying the exclusion of a charge-injection barrier.

According to the discussions above, band-to-band tunneling[28] is expected to be a dominant cause of the superlinearity observed in BLG FETs. Tunneling currents have a much weaker temperature dependence than thermionic-emission currents. To measure the temperature dependence, as-fabricated BLG FETs were introduced into a vacuum probe station and the $I_D$–$V_D$ characteristics were measured at $V_{NP}$ in the temperature range from 83 to 280 K. The pressure of the system during the measurements was on the order of $10^{-4}$ Pa. A FET with $L = 0.2$ μm was used because FETs with shorter channels have larger contributions from the metal contacts. There were almost no dependences of the degree of nonlinearity, $\Delta(dI_D/dV_D)$, and the zero-bias conductance on the temperature, as shown in Fig. 3a and b, respectively. These results support the above theory where the superlinear curve in back-gate BLG FETs is caused by tunneling through the bandgap.

When a graphene layer is hole-doped by the contact, the hole concentration in the upper graphene layer that is in direct contact with the metallic electrode is higher than that of the lower layer owing to screening.[29–31] The charge screening length has been reported to be 1.02 in graphene.[31] Thus, the hole concentration in the lower layer is approximately 1/$e$ of the upper layer. The potential difference caused by the screening induces a bandgap in BLG at the contacts. The magnitude of the bandgap is determined by electrical displacement fields. The displacement field ($\bar{D}$ in Ref. 8) corresponding to the $\Delta E_F$ of 0.25 eV (Fig. 2b) is 0.62 V nm$^{-1}$, which should induce a bandgap of roughly 60 meV at $V_G = V_{NP}$.[8] Figure 4 shows the expected band diagrams for the three $V_G$ regions. Partial pinning of the charge density of BLG under the contacts is possibly allowed in actual systems,[32,33] and thus the bandgap should be slightly tuned by $V_G$ from 60 meV. Considering the hole doping from the contacts to BLG, positive and negative $V_G$'s increase and decrease the bandgap, respectively.[8] A tunneling current should be observed when $V_G > V_{NP}$ through two $pn$ junctions formed along the channel, and a superlinearity should be observed in the same $V_G$ range. However, a superlinearity was also observed around $V_{NP}$ as shown in Fig. 1d.

To further characterize the $V_G$ dependence of the nonlinearity, four-terminal measurements were performed using a BLG flake patterned by oxygen plasma etching (Fig. 5a). The inner potential probes were 1.6 μm from the channel, guaranteeing an invasive nature of the probes.[34,35] The device was annealed under vacuum (~10$^{-3}$ Pa) at 400 K for 20 h and characterized without breaking the vacuum. The four-terminal data also showed a superlinearity around $V_{NP}$ (Fig. 5b). Thus, the superlinearity around $V_{NP}$ is not characteristic of the metal contact, but of the channel. This feature might be a result of electron and hole puddles in the channel.[36,37] A potential fluctuation caused by the puddles could induce a bandgap, and as a result, tunneling conduction should be unavoidable around $V_{NP}$. The superlinearity consistently decreased as the absolute gate voltage increased. In contrast, the two-terminal data (using the outer electrodes; Fig. 5c) showed clear asymmetry with respect to $V_{NP}$. As observed with the as-fabricated BLG FETs (Fig. 1d), the degree of the superlinearity somewhat increased in the positively-gated region as $V_G$ increased. This feature can be attributed to tunneling conduction through the bandgap at the $pn$ junctions formed by the metal contacts (Fig. 4c).

Finally, the electrode contact resistances ($R_C$) were extracted using the transfer length method from the channel-length dependent transfer characteristics shown in Fig. 2a. To avoid artifacts, $R_C$ was calculated using $V_G - V_{NP}$ instead of $V_G$.[38] The extracted $R_C$ values are plotted in Fig. 2c. The slight decrease around $V_{NP}$ in the $R_C$–$V_G$ characteristics was caused by carrier doping from the contacts and the higher $R_C$ in the positively-gated region is indicative of hole doping from the contacts.[33] The degree of nonlinearity, $\Delta(dI_D/dV_D)$, was higher in the positively-gated region, i.e., in the electron conduction branch, as shown in Figs. 1d and 5c. Thus, the higher $R_C$ in the positively-gated region is partly attributable to an additional resistance caused by opening the bandgap and the resultant tunneling resistances.

In conclusion, bandgap opening in BLG at the metal contacts was investigated using electrical characterization. Superlinear current-voltage characteristics caused by tunneling conduction through the bandgap were observed in BLG FETs. The observed superlinearity was higher in the positively-gated regions ($V_G > V_{NP}$), which is consistent with hole doping from the contacts (Cr/Au electrodes were used) and band-to-band tunneling through a resultant *pnp* junction. BLG is considered to be a promising material for use in digital applications owing to its (tunable) bandgap. Application in electronic devices usually requires a two-terminal device configuration with a total resistance containing the electrode-contact resistances. Bandgap opening at the metal-BLG interface is an additional source of contact resistance and should be carefully considered in future applications.

This work was supported in part by the Special Coordination Funds for Promoting Science and Technology, and a Grant-in-Aid for Scientific Research on Innovative Areas (No. 26107531) from the Ministry of Education, Culture, Sports, Science and Technology of Japan; and by a technology research grant from the JFE 21st Century Foundation. The graphite crystal used was supplied by M. Murakami and M. Shiraishi.

1. S. V. Morozov, K. S. Novoselov, M. I. Katsnelson, F. Schedin, D. C. Elias, J. A. Jaszczak, and A. K. Geim, Phys. Rev. Lett. **100**, 016602 (2008).

2. Y.-W. Son, M. L. Cohen, and S. G. Louie, Phys. Rev. Lett. **97**, 216803 (2006).

3. M. Y. Han, B. Özyilmaz, Y. B. Zhang, and P. Kim, Phys. Rev. Lett. **98**, 206805 (2007).

4. Z. Chen, Y.-M. Lin, M. J. Rooks, and Ph. Avouris, Physica E **40**, 228 (2007).

5. H. Min, B. Sahu, S. K. Banerjee, and A. H. MacDonald, Phys. Rev. B **75**, 155115 (2007).

6. J. B. Oostinga, H. B. Heersche, X. L. Liu, A. F. Morpurgo, and L. M. K. Vandersypen, Nat. Mater. **7**, 151 (2008).

7. E. V. Castro, K. S. Novoselov, S. V. Morozov, N. M. R. Peres, J. M. B. Lopes dos Santos, J. Nilsson, F. Guinea, A. K. Geim, and A. H. Castro Neto, Phys. Rev. Lett. **99**, 216802 (2007).

8. Y. Zhang, T.-T. Tang, C. Girit, Z. Hao, M. C. Martin, A. Zettl, M. F. Crommie, Y. R. Shen, and F. Wang, Nature **459**, 820 (2009).

9. C. Berger, Z. Song, X. Li, X. Wu, N. Brown, C. Naud, D. Mayou, T. Li, J. Hass, A. N. Marchenkov, E. H. Conrad, P. N. First, and W. A. de Heer, Science **312**, *312*, 1191 (2006).

10. D. Gunlycke, D. A. Areshkin, and C. T. White, Appl. Phys. Lett. **90**, 142104 (2007).

11. T. Ohta, A. Bostwick, T. Seyller, K. Horn, and E. Rotenberg, Science **313**, 951 (2006).

12. J. Park, S. B. Jo, Y.-J. Yu, Y. Kim, J. W. Yang, W. H. Lee, H. H. Kim, B. H. Hong, P. Kim, K. Cho, and K. S. Kim, Adv. Mater. **24**, 407 (2012).

13. S. Ryu, L. Liu, S. Berciaud, Y.-J. Yu, H. Liu, P. Kim, G. W. Flynn, and L. E. Brus, Nano Lett. **10**, 4944 (2010).

14. J. Moser, A. Verdaguer, D. Jiménez, A. Barreiro, and A. Bachtold, Appl. Phys. Lett. **92**, 123507 (2008).

15. I. Meric, M. Y. Han, A. F. Young, B. Özyilmaz, P. Kim, and K. L. Shepard, Nat. Nanotechnol. **3**, 654 (2008).

16. A. Barreiro, M. Lazzeri, J. Moser, F. Mauri, and A. Bachtold, Phys. Rev. Lett. **103**, 076601 (2009).

17. V. Perebeinos and Ph. Avouris, Phys. Rev. B **81**, 195442 (2010).

18. S. M. Sze and K. K. Ng, *Physics of Semiconductor Devices*, 3rd ed. (Wiley, New York, 2007).

19. J. G. Simmons, *J. Appl. Phys.* **34**, 2581 (1963).

20. R. Nouchi and K. Tanigaki, Appl. Phys. Lett. **96**, 253503 (2010).

21. R. Nouchi and K. Tanigaki, Appl. Phys. Lett. **105**, 033112 (2014).

22. A. Varykhalov, M. R. Scholz, T. K. Kim, and O. Rader, Phys. Rev. B **82**, 121101(R) (2010).

23. J. Zheng, Y. Wang, L. Wang, R. Quhe, Z. Ni, W.-N. Mei, Z. Gao, D. Yu, J. Shi, and J. Lu, Sci. Rep. **3**, 2081 (2013).

24. R. Nouchi, T. Saito, and K. Tanigaki, Appl. Phys. Express **4**, 035101 (2011).

25. T. Mueller, F. Xia, M. Freitag, J. Tsang, and Ph. Avouris, Phys. Rev. B **79**, 245430 (2009).

26. Y.-J. Yu, Y. Zhao, S. Ryu, L. E. Brus, K. S. Kim, and P. Kim, Nano Lett. **9**, 3430 (2009).

27. L. M. Zhang, Z. Q. Li, D. N. Basov, and M. M. Fogler, Z. Hao, and M. C. Martin, Phys. Rev. B **78**, 235408 (2008).

28. H. Miyazaki, M. V. Lee, S.-L. Li, H. Hiura, A. Kanda, and K. Tsukagoshi, J. Phys. Soc. Jpn. **81**, 014708 (2012).

29. T. Ohta, A. Bostwick, J. L. McChesney, T. Seyller, K. Horn, and E. Rotenberg, Phys. Rev. Lett. **98**, 206802 (2007).

30. H. Miyazaki, S. Odaka, T. Sato, S. Tanaka, H. Goto, A. Kanda, K. Tsukagoshi, Y. Ootuka, and Y. Aoyagi, Appl. Phys. Express **1**, 034007 (2008).

31. D. Sun, C. Divin, C. Berger, W. A. de Heer, P. N. First, and T. B. Norris, Phys. Rev. Lett. **104**, 136802 (2010).

32. A. Di Bartolomeo, F. Giubileo, S. Santandrea, F. Romeo, R. Citro, T. Schroeder, and G. Lupina, Nanotechnology **22**, 275702 (2011).

33. R. Nouchi, T. Saito, and K. Tanigaki, J. Appl. Phys. **111**, 084314 (2012).

34. B. Huard, N. Stander, J. A. Sulpizio, and D. Goldhaber-Gordon, Phys. Rev. B **78**, 121402(R) (2008).

35. R. Golizadeh-Mojarad and S. Datta, Phys. Rev. B **79**, 085410 (2009).

36. E. H. Hwang, S. Adam, and S. Das Sarma, Phys. Rev. Lett. **98**, 186806 (2007).

37. J. Martin, N. Akerman, G. Ulbricht, T. Lohmann, J. H. Smet, K. von Klitzing, and A. Yacoby, Nat. Phys. **4**, 144 (2008).

38. K. Nagashio and A. Toriumi, Jpn. J. Appl. Phys. **50**, 070108 (2011).

**Figures**

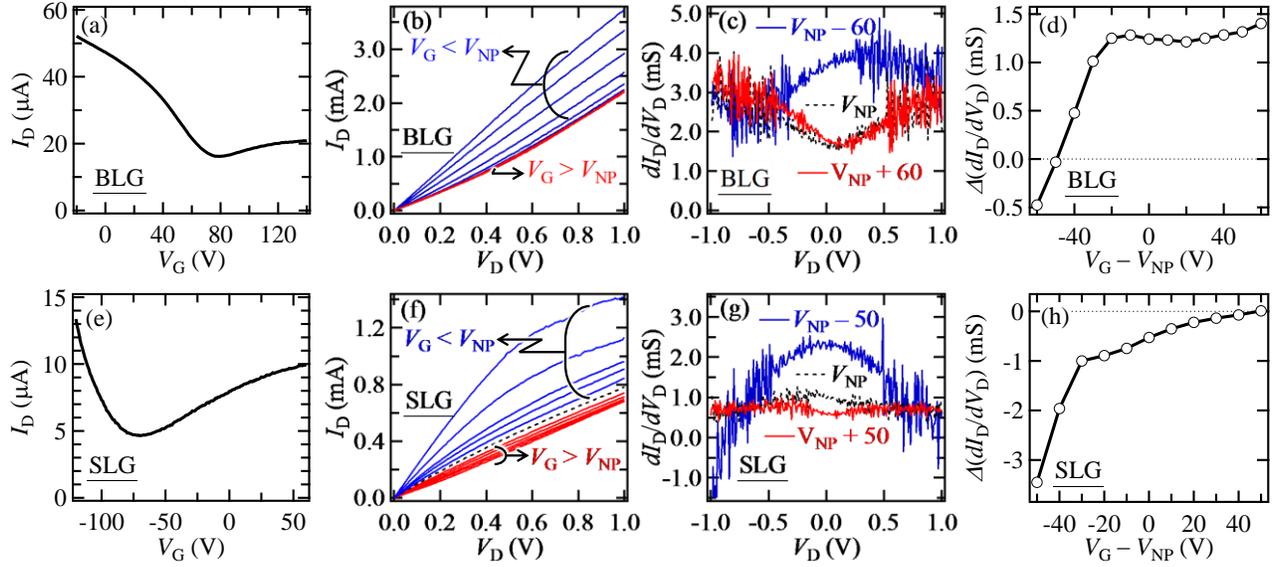

FIG. 1. Electrical properties of (a-d) BLG and (e-h) SLG FETs measured in air in the dark at room temperature with $L = 0.5$ μm. (a,e) Transfer characteristics measured with $V_D = 10$ mV. (b,f) Output characteristics measured within $V_{NP} - 60$ (V) $\leq V_G \leq V_{NP} + 60$ (V) for BLG, and $V_{NP} - 50$ (V) $\leq V_G \leq V_{NP} + 50$ (V) for SLG. (c,g) First-order derivative of $I_D$, $dI_D/dV_D$, numerically calculated from the data in (b,f). (d,h) Differences in the derivatives at 1 V and 0 V, $\Delta(dI_D/dV_D) \equiv dI_D/dV_D|_{1\ V} - dI_D/dV_D|_{0\ V}$, obtained by linearly fitting the data in (c,g).

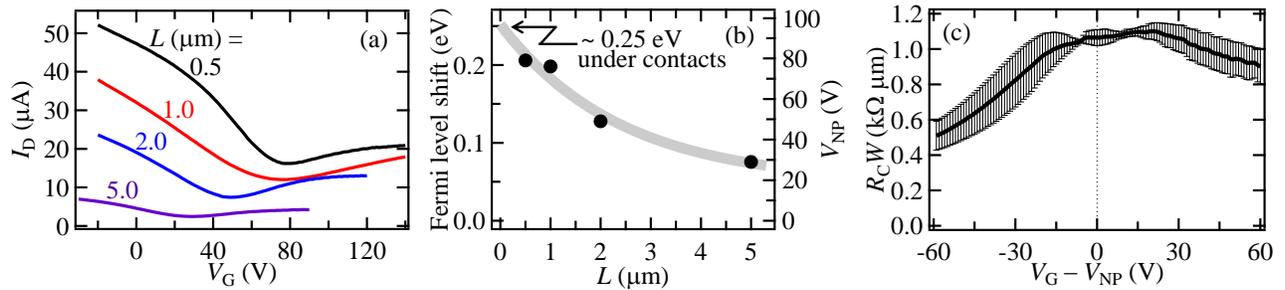

FIG. 2. Channel-length dependence of the BLG FETs measured in air in the dark and at room temperature. (a) Transfer characteristics measured with $V_D = 10$ mV. (b) The charge neutrality point, $V_{NP}$, and corresponding shifts of the Fermi level. The Fermi level shift of BLG under metal contacts was estimated to be 0.25 eV by extrapolation. (c) Contact resistance, numerically obtained from the data in (a) using the transfer length method. The error bars are standard errors of the least squares fit.

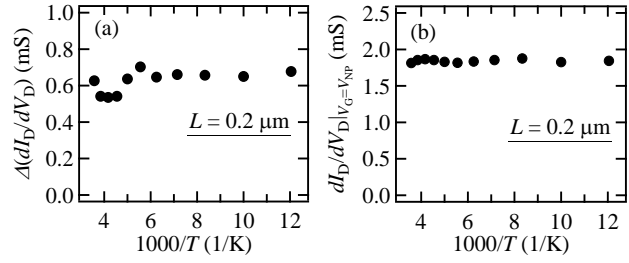

FIG. 3. Temperature dependence of (a) $\Delta(dI_D/dV_D)$ and (b) the zero-bias conductance of a BLG FET measured under a high vacuum in the dark with $L = 0.2$ μm and $V_G = V_{NP}$.

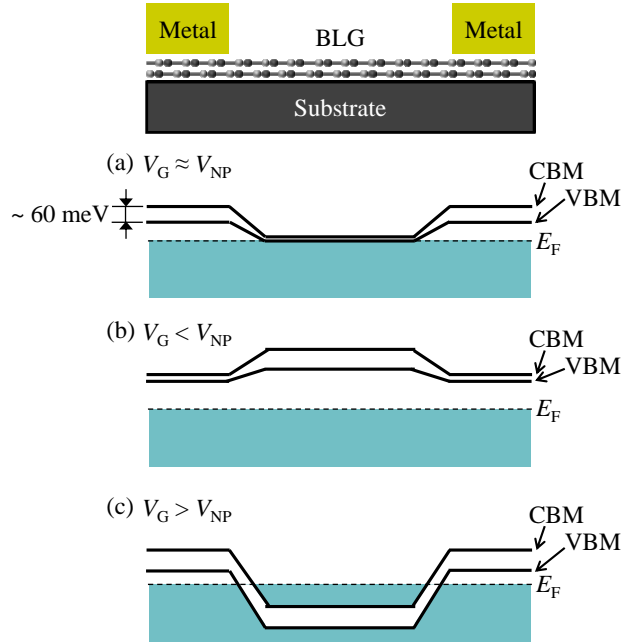

FIG. 4. Expected band diagrams for a BLG FET. The three $V_G$ regions, (a) $V_G \approx V_{NP}$, (b) $V_G < V_{NP}$ and (c) $V_G > V_{NP}$ are displayed. CBM, VBM and $E_F$ stand for the conduction band minimum, the valence band maximum and the Fermi level, respectively. The bandgap at $V_{NP}$ was estimated to be 60 meV from the data in Fig. 2b.

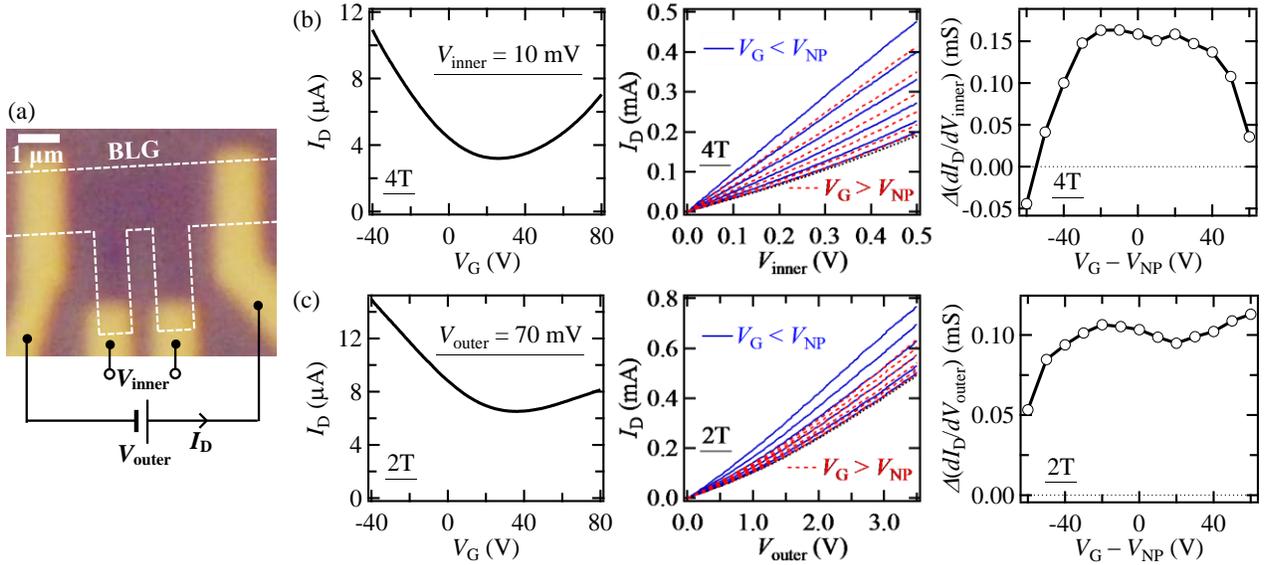

FIG. 5. (a) Optical micrograph of a BLG FET with a four-terminal configuration. The BLG flake is outlined by the white dashed line. Electrical properties acquired by (b) four-terminal (4T) measurements using noninvasive potential probes with a separation of 0.5 μm and (c) two-terminal (2T) measurements using invasive outer electrodes with a separation of 3.5 μm. The electric field strength was equalized between these two measurements. The transfer characteristics were measured with $V_{inner}$ = 10 mV and $V_{outer}$ = 70 mV for the 4T and 2T measurements. The output characteristics were measured within $V_{NP}$ − 60 (V) ≤ $V_G$ ≤ $V_{NP}$ + 60 (V). The degree of the nonlinearity, the difference in the first-order derivative of $I_D$ at the maximum $V_D$ (0.5 and 3.5 V for 4T and 2T measurements, respectively) and 0 V, was numerically obtained from the output characteristics.